\documentclass{article}

\usepackage[latin1]{inputenc}
\usepackage{times}
\usepackage{latexsym}
\usepackage{amsmath}
\usepackage{amssymb}
\usepackage{graphicx}
\usepackage{fullpage}
\usepackage{ifthen}

\newcommand{\nc}{\newcommand}
\nc{\rnc}{\renewcommand}
\nc{\nev}{\newenvironment}

\rnc{\subsection}{\secdef\ssa\ssb}
\nc{\ssa}[2][default]{\par\vspace{2ex}\refstepcounter{subsection}\noindent\textbf{\thesubsection. #1. }}
\nc{\ssb}[1]{\par\bigskip\noindent\textbf{#1. }}

\makeatletter
\rnc{\@seccntformat}[1]{{\normalfont\bfseries{\csname the#1\endcsname}\hspace{1pt}.\hspace{0.4em}}}
\rnc{\section}{\@startsection
        {section}%
        {1}%
        {0mm}%
        {-\baselineskip}%
        {0.5\baselineskip}%
        {\normalfont\normalsize\bfseries\centering}%
}
\renewcommand{\@makecaption}[2]{\begin{center}#1. #2\end{center}}
\makeatother

\newcounter{theo}[section]
\rnc{\thetheo}{\thesection.\arabic{theo}}
 
\nev{defform}[2]{\refstepcounter{theo}\begin{list}{}{%
      \setlength{\labelwidth}{0em}%
      \setlength{\leftmargin}{0em}%
      \setlength{\itemindent}{\labelsep}%
      \setlength{\listparindent}{1.5em}
      \setlength{\parsep}{0em}}%
      \item[\textbf{#2 \thetheo\ifthenelse{\equal{#1}{}}{}{ (#1)}.}]}%
    {\end{list}}

\nev{satzform}[2]{\begin{defform}{#1}{#2}\itshape}{\end{defform}}

\nev{theo}[1][]{\begin{satzform}{#1}{Theorem}}{\end{satzform}}
\nev{lem}[1][]{\begin{satzform}{#1}{Lemma}}{\end{satzform}}
\nev{cor}[1][]{\begin{satzform}{#1}{Corollary}}{\end{satzform}}
\nev{prop}[1][]{\begin{satzform}{#1}{Proposition}{}}{\end{satzform}}
\nev{fact}[1][]{\begin{satzform}{#1}{Fact}}{\end{satzform}}
\nev{exa}[1][]{\begin{defform}{#1}{Example}}{\end{defform}}
\nev{defn}[1][]{\begin{defform}{#1}{Definition}}{\end{defform}}
\nev{rem}[1][]{\begin{defform}{#1}{Remark}}{\end{defform}}

\nc{\proof}{\medskip\noindent\textit{Proof: }}
\nc{\proofend}{\hfill$\Box$\vspace{\topsep}\par}

\rnc{\labelenumi}{(\arabic{enumi})}
\rnc{\labelitemi}{\text{--}}
\rnc{\phi}{\varphi}

\newlength{\probwidth}
\nc{\prob}[3][9]{\begin{center}\normalfont\fbox{\begin{tabular}[t]{rp{#1cm}}\textit{Input:}&#2.\\\textit{Problem:}&#3.\end{tabular}}\end{center}}

\begin{document}
\title{\large\bfseries Computing Crossing Numbers in Quadratic Time}
\author{\normalsize Martin Grohe\\
\normalsize University of Illinois at Chicago}
\date{\normalsize\today}
\maketitle

{\renewcommand{\thefootnote}{}\footnotetext{Author's address: 
Martin Grohe, Department of Mathematics, Statistics, and Computer Science,
University of Illinois at Chicago, 851 S.~Morgan St.\ (M/C 249),
Chicago, IL~60607-7045, USA. Email: grohe@math.uic.edu.}}

\begin{abstract}
  We show that for every fixed $k\ge 0$ there is a quadratic time
  algorithm that decides whether a given graph has crossing number at
  most $k$ and, if this is the case, computes a drawing of the graph
  in the plane with at most $k$ crossings.
 \end{abstract}

\section{Introduction}
Hopcroft and Tarjan \cite{hoptar74} showed in 1974 that planarity of
graphs can be decided in linear time. It is natural to relax planarity
by admitting a small number of edge-crossings in a drawing of the
graph. The \emph{crossing number} of a graph is the minimum number of
edge crossings needed in a drawing of the graph in the plane. Not
surprisingly, it is NP-complete to decide, given a graph $G$ and a $k$,
whether the crossing number of $G$ is at most $k$ \cite{garjoh82}. On
the other hand, for every \emph{fixed} $k$ there is a simple
polynomial time algorithm deciding whether a given graph $G$ has
crossing number at most $k$: It guesses $l\le k$ pairs of edges that
cross\footnote{This can be implemented by exhaustive search of the
  space of $m^{2k}$ $k$-tuples of edge pairs, where $m$ denotes the
  number of edges of the input graph.} and tests if the graph obtained
from $G$ by adding a new vertex at each of these edge crossings is
planar. The running time of this algorithm is $n^{\Theta(k)}$. Downey
and Fellows \cite{dowfel99} raised the question
if the crossing-number problem is \emph{fixed parameter-tractable},
that is, if there is a constant $c\ge 1$ such that for every fixed
$k$ the problem can be solved in time $O(n^c)$. We answer this
question positively with $c=2$. In other words, we show that for every
fixed $k$ there is a quadratic time algorithm deciding whether a given
graph $G$ has crossing number at most $k$. Moreover, we show that if
this is the case, a drawing of $G$ in the plane with at most $k$
crossings can also be computed in quadratic time.

It is interesting to compare our result to similar results for
computing the \emph{genus} of a graph. (The genus of a graph $G$ is
the minimum taken over the genus of all surfaces $S$ such that $G$ can
be embedded into $S$.)  As for the crossing number, it is NP-complete
to decide if the genus of a given graph is less than or equal to a
given $k$ \cite{tho88}. For a fixed $k$, at first sight the genus problem looks much
harder. It is by no means obvious how to solve it in polynomial time;
this has been proved possible by Filotti, Miller, and Reif
\cite{filmilrei79}. In 1996, Mohar \cite{moh96} proved that for every
$k$ there is actually a linear time algorithm deciding whether the
genus of a given graph is $k$.  However, the fact that the genus
problem is fixed-parameter tractable was known earlier as a direct
consequence of a strong general theorem due to Robertson and Seymour
\cite{GMXIII} stating that all minor closed classes of graphs are
recognizable in cubic time. It is easy to see that the class of graphs
of genus at most $k$ is closed under taking minors, but unfortunately
the class of all graphs of crossing number at most $k$ is not. So in
general Robertson and Seymour's theorem cannot be applied to compute
crossing numbers. An exception is the case of graphs of degree at
most 3; Fellows and Langston \cite{fellan88} observed that for such
graphs Robertson and Seymour's result immediately yields a
cubic time algorithm for computing crossing numbers.\footnote{This is
  simply because for graphs of degree at most 3 the minor relation and
  the topological subgraph relation coincide.}

Although we cannot apply Robertson and Seymour's result directly, the
overall strategy of our algorithm is inspired by their
ideas: The algorithm first iteratively reduces the size of the input
graph until it reaches a graph of bounded tree-width, and then solves
the problem on this graph. For the reduction step, we use Robertson
and Seymour's Excluded Grid Theorem \cite{GMV} together with a nice
observation due to Thomassen \cite{tho97} that in a graph of bounded
genus (and thus in a graph of bounded crossing number) every large
grid contains a subgrid that, in some precise sense, lies ``flat'' in
the graph. Such a flat grid does not essentially contribute to the
crossing number and can therefore be contracted.  For the remaining
problem on graphs of bounded tree-width we apply a theorem due to
Courcelle \cite{cou90} stating that all properties of graphs that are
expressible in monadic second-order logic are decidable in linear time
on graphs of bounded tree-width.

Let me remark that the hidden constant in the quadratic upper bound
for the running time of our algorithm
heavily depends on $k$. As a matter of fact, the running time is
$O(f(k)\cdot n^2)$, where $f$ is a doubly exponential function. Thus
our algorithm is mainly of theoretical interest.

\section{Preliminaries}
Graphs in this paper are undirected and loop-free, but they may have multiple
edges.\footnote{Note that loops are completely irrelevant for the crossing
  number, whereas multiple edges are not.} The vertex set of a graph $G$ is
denoted by $V^G$, the edge set by $E^G$. For graphs $G$
and $H$ we let $G\cup H:=(V^G\cup V^H,E^G\cup E^H)$ and $G\setminus
H:=\big(V^G\setminus V^H,\{e\in E^G\setminus E^H\mid\text{both endpoints of $e$
  are contained in $V^G\setminus V^H$}\}\big)$.

\subsection{Topological Embeddings}
A \emph{topological embedding} of a graph $G$ into a graph $H$ is a mapping $h$
that associates a vertex $h(v)\in V^H$ with every $v\in V^G$ and a path $h(e)$
in $H$ with every $e\in E^G$ in such a way that:
\begin{itemize}
\item For distinct vertices $v,w\in V^G$, the vertices $h(v)$ and $h(w)$
  are distinct.
\item For distinct edges $e,f\in E^G$, the paths $h(e)$ and $h(f)$ are
  internally disjoint (that is, they have at most their endpoints in common).
\item For every edge $e\in E^G$ with endpoints $v$ and $w$, the two endpoints
  of the path $h(e)$ are $h(v)$ and $h(w)$, and $h(u)\not\in V^{h(e)}$ for all
  $u\in V^G\setminus\{v,w\}$.
\end{itemize}
We let $h(G):=\big(h(V^G),\emptyset\big)\cup\bigcup_{e\in E^G}h(e)$.

\subsection{Drawings and Crossing Numbers}
A \emph{drawing} of a graph $G$ is a mapping $\Delta$ that associates with
every vertex $v\in V^G$ a point $\Delta(v)\in\mathbb R^2$ and with every edge
$e\in E^G$ a simple curve $\Delta(e)$ in $\mathbb R^2$ in such a way that:
\begin{itemize}
\item For distinct vertices $v,w\in V^G$, the points $\Delta(v)$ and $\Delta(w)$
  are distinct.
\item
For distinct edges $e,f\in E^G$, the curves $\Delta(e)$ and $\Delta(f)$
have at most one interior point in common (and possibly their endpoints).
\item For every edge $e\in E^G$ with endpoints $v$ and $w$, the two endpoints
  of the curve $\Delta(e)$ are $\Delta(v)$ and $\Delta(w)$, and
  $\Delta(u)\not\in\Delta(e)$ for all $u\in V^G\setminus\{v,w\}$.
\item
At most two edges intersect in one point. More precisely, $|\{e\in E^G\mid x\in\Delta(e)\}|\le 2$ for all $x\in\mathbb R^2\setminus\Delta(V^G)$.
\end{itemize}
We let $\Delta(G):=\Delta(V^G)\cup\bigcup_{e\in E^G}\Delta(e)$.

An $x\in\mathbb R^2\setminus\Delta(V^G)$ with $|\{e\in E^G\mid
x\in\Delta(e)\}|=2$ is called a \emph{crossing} of $\Delta$. The \emph{crossing
number} of $\Delta$ is the number of crossings of $\Delta$. The \emph{crossing 
number} of $G$ is the minimum taken over the crossing numbers of all drawings
of $G$. A drawing or graph of crossing number 0 is called \emph{planar}.

\subsection{Hexagonal Grids}
\begin{figure}[ht]
\mbox{}
\hfill
\includegraphics[width=.8cm]{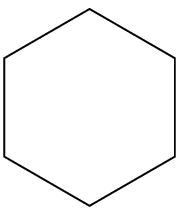}
\hfill
\includegraphics[width=2.4cm]{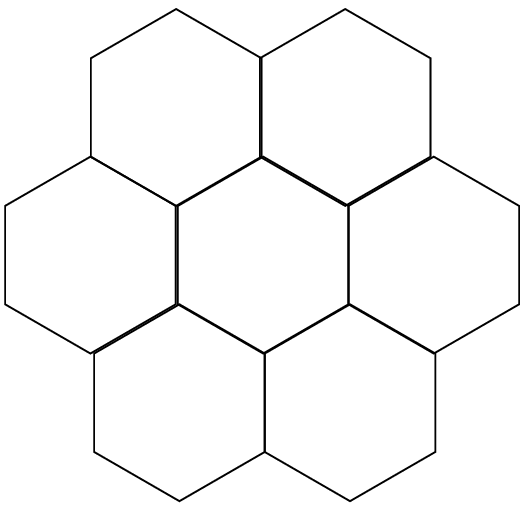}
\hfill
\includegraphics[width=4cm]{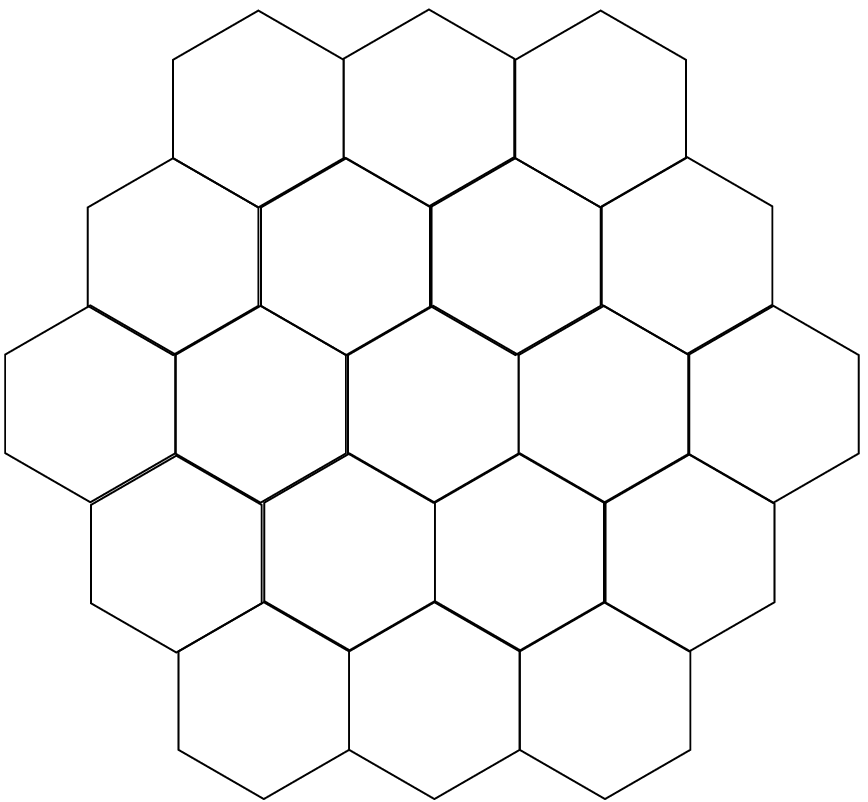}
\hfill
\mbox{}
\caption{The hexagonal grids $H_1,H_2,H_3$}\label{fig:hex}
\end{figure}

\begin{figure}[ht]
\centering
\includegraphics[width=4cm]{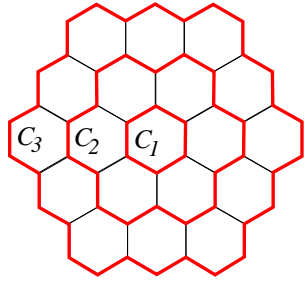}
\caption{The principal cycles of $H_3$}\label{fig:pricyc}
\end{figure}

For $r\ge 1$, we let $H_r$ be the hexagonal grid of radius $r$. Instead of
giving a formal definition, we refer the reader to Figure \ref{fig:hex} to see
what this means. 
The \emph{principal cycles} $C_1,\ldots,C_r$ of $H_r$ are the
the concentric cycles, numbered from the interior to the exterior (see Figure \ref{fig:pricyc}).  

\subsection{Flat Grids in a Graph}
For graphs $H\subseteq G$, an \emph{$H$-component (of $G$)} is either a
connected component $C$ of $G\setminus H$ together with all edges
connecting $C$ with $H$ and their endpoints in $H$ or an edge in
$E^G\setminus E^H$ whose endpoints are both in $H$ together with its
endpoints.  Let $G$ be a graph and $h:H_r\to G$ a topological embedding. The
\emph{interior of $h(H_r)$} is the subgraph $h(H_r\setminus C_r)$
(remember that $C_r$ is the outermost principal cycle of $H_r$). The
\emph{attachments of $h(H_r)$} are those $h(H_r)$-components that have
a non-empty intersection with the interior of $h(H_r)$. The
topological embedding $h$ is \emph{flat} if the union of $h(H_r)$ with all its
attachments is planar.

We shall use the following theorem due to Thomassen \cite{tho97}. Actually,
Thomassen stated the result for the \emph{genus} of a graph rather than its
crossing number. However, it is easy to see that the crossing number of a
graph is an upper bound for its genus.

\begin{theo}[Thomassen \cite{tho97}]\label{theo:tho}
For all $k,r\ge 1$ there is an $s\ge 1$ such that the following holds: 
If $G$ is a graph of crossing number at most $k$ and $h:H_s\to G$ a
topological embedding, then there is a subgrid $H_r\subseteq H_s$ such that the
restriction $h|_{H_r}$ of $h$ to $H_r$ is flat.
\end{theo}

\subsection{Tree-Width}
We assume that reader is familiar with the notion \emph{tree-width (of a
  graph)}. It is no big problem if not; we never really work with
tree-width, but just take it as a black box in Theorems
\ref{theo:rsb}--\ref{theo:cou2}. Robertson and Seymour's deep
\emph{Excluded Grid Theorem} \cite{GMV} states that every graph of
sufficiently large tree-width contains the homeomorphic image of a large grid.
The following is an algorithmic version of this theorem.

\begin{theo}[Robertson, Seymour \cite{GMXIII}, Bodlaender \cite{bod96}]\label{theo:rsb}
  Let $r\ge 1$. Then there is a $w\ge 1$ and a linear time algorithm that,
  given a graph $G$, either (correctly) recognizes that the tree-width of $G$
  is at most $w$ or computes a topological embedding $h:H_r\to G$.
\end{theo}

Actually, in \cite{GMXIII} Robertson and Seymour only give a quadratic time
algorithm, but they point out that their algorithm can be improved to linear
time using Bodlaender's \cite{bod96} linear time algorithm for computing
tree-decompositions. Let me remark that, as far as I can see, this
algorithm is not merely a trivial modification of Robertson and Seymour's
algorithm obtained by ``plugging in'' Bodlaender's tree-decomposition
algorithm, but it requires to look into the details of Bodlaender's algorithm
and extend it in a suitable way. 

\subsection{Courcelle's Theorem}
Courcelle's theorem states that properties of graphs definable in
\emph{Monadic Second-Order Logic MSO} can be checked in linear time.
In this logical context we consider graphs as relational structures of
vocabulary $\{E,V,I\}$, where $V$ and $E$ are unary relation symbols
interpreted as the vertex set and edge set, respectively, and $I$ is a
binary relation symbol interpreted by the incidence relation of a
graph. To simplify the notation, for a graph $G$ we let $U^G:=V^G\cup
E^G$ and call $U^G$ the \emph{universe} of $G$.

I assume that the reader is familiar with the definition of MSO. However, for
those who are not I have included it in Appendix A.

\begin{theo}[Courcelle \cite{cou90}]\label{theo:cou}
  Let $w\ge 1$ and let $\phi(x_1,\ldots,x_k,X_1,\ldots,X_l)$ be an
  MSO-formula. Then there is a linear time algorithm that, given a graph $G$
  and $a_1,\ldots,a_k\in U^G$, $A_1,\ldots,A_l\subseteq U^G$,
  decides whether $G\models\phi(a_1,\ldots,a_k,A_1,\ldots,A_l)$.
\end{theo}

We shall also use the following strengthening of Courcelle's
theorem, a proof of which can be found in \cite{flufrigro01}:

\begin{theo}\label{theo:cou2}
  Let $w\ge 1$ and let
  $\phi(x_1,\ldots,x_k,X_1,\ldots,X_l,y_1,\ldots,y_m,Y_1,\ldots,Y_n)$ be an
  MSO-formu\-la. Then there is a linear time algorithm that, given a graph $G$
  and $b_1,\ldots,b_m\in U^G$, $B_1,\ldots,B_n\subseteq U^G$,
  decides if there exist $a_1,\ldots,a_k\in U^G$,
  $A_1,\ldots,A_l\subseteq U^G$ such that
  \[
G\models\phi(a_1,\ldots,a_k,A_1,\ldots,A_l,b_1,\ldots,b_m,B_1,\ldots,B_n),
\]
  and, if this is the case, computes such elements $a_1,\ldots,a_k$ and sets $A_1,\ldots,A_l$.
\end{theo}

\section{The Algorithm}
For an $l\ge 1$, a graph $G$, and a subset
$F\subseteq E^G$ of \emph{forbidden edges}, an \emph{$l$-good drawing of
$G$ with respect to $F$} is a drawing $\Delta$ of $G$ of crossing
number at most $l$ such that no forbidden edges are involved in any
crossings, i.e.\ for every crossing $x\in\Delta(e)\cap\Delta(f)$ of
$\Delta$ we have $e,f\in E^G\setminus F$.

We fix a $k\ge 1$ for the whole section. We shall describe an algorithm that
solves the following \emph{generalized $k$-crossing number problem} in
quadratic time:

\prob{Graph $G$ and subset $F\subseteq E^G$}{Decide if $G$ has a $k$-good
  drawing with respect to $F$}

Later, we shall extend our algorithm in such a way that it actually computes a
$k$-good drawing if there exists one.

Our algorithm works in two phases. In the first, it iteratively reduces the
size of the input graph until it obtains a graph whose tree-width is
bounded by a constant only depending on $k$. Then, in the second phase, it
solves the problem on this graph of bounded tree-width.

\subsection*{Phase I}
We let $r:=2k+2$ and choose $s$ sufficiently large such that for every graph
$G$ of crossing number at most $k$  and every topological embedding $h:H_s\to G$ there is a subgrid $H_r\subseteq H_s$ such that the
restriction $h|_{H_r}$ of $h$ to $H_r$ is flat. Such an $s$ exists by Theorem
\ref{theo:tho}. Then we choose $w$ with respect to $s$ according
to Theorem \ref{theo:rsb} such that we have a linear time algorithm that, given a
graph of tree-width at least $w$,  finds a topological embedding $h:H_s\to G$.
We keep $r,s,w$ fixed for the rest of the section.

\begin{lem}\label{lem:ml1}
  There is a linear time algorithm that, given a graph $G$, either recognizes
  that the crossing number of $G$ is greater than $k$, or recognizes that the
  tree-width of $G$ is at most $w$, or computes a flat topological embedding $h:H_r\to
  G$.
\end{lem}

\proof We first apply the algorithm of Theorem \ref{theo:rsb}. If it
recognizes that the tree-width of the input graph $G$ is at most $w$,
we are done. Otherwise, it computes a topological embedding $h:H_s\to
G$. By our choice of $s$, we know that either the crossing number of
$G$ is greater than $k$ or there is a subgrid $H_r\subseteq H_s$ such
that the restriction of $h$ to $H_r$ is flat.

For each $H_r\subseteq H_s$ we can decide whether $h|_{H_r}$ is
flat by a planarity test, which is possible in linear time
\cite{hoptar74}. Our algorithm tests whether $h|_{H_r}$ is
flat for all $H_r\subseteq H_s$. Either it finds a flat $h|_{H_r}$, or the
crossing number of $G$ is greater than $k$.\footnote{A look at the proof of
  Thomassens's theorem reveals that we do not have to test all $H_r\subseteq
  H_s$ for flatness, but only a number that is linear in $k$.}

Since $s$ is a fixed constant, the overall running time is linear.
\proofend

Let $G$ be a graph and $h:H_r\to G$ a flat topological embedding. For $2\le
i\le r$, we let $H^i$ be the subgrid of $H_r$ bounded by the $i$th principal
cycle $C_i$. We let $K_i$ be the subgraph of $G$ consisting of $h(H^i)$ and
all attachments of $h(H_r)$ intersecting the interior
$h(H^i\setminus C_{i})$ of $h(H_i)$. Moreover, we let $F_i$ be the set of all
edges of $K_i$ that have at least one endpoint on $h(C_i)$. Using the fact
that $h$ is flat, it is easy to see that the sets $F_i$, for $2\le i\le r$ are
disjoint.

Suppose now that $\Delta$ is a $k$-good drawing of $G$ of minimum crossing
number. Recall that $r=2k+2$. By the pigeonhole-principle there is at least
one $i, 2\le i\le r$ such that none of the edges in $F_i$ is involved in any
crossing of $\Delta$. We let $i_0, 2\le i\le r$ be minimum with this property.

Let $C:=h(C_{i_0})$, $K:=K_{i_0}$ and $I:=K\setminus C$. Then $K$ and $I$ are
both connected planar graphs. Note furthermore that $\Delta(C)$ is a simple
closed curve in the plane $\mathbb R^2$. Thus $\Delta(I)$ must be entirely
contained in one connected component of $\mathbb R^2\setminus \Delta(C)$, say,
in the interior.

I claim that the restriction of $\Delta$ to $K$ is a planar drawing. Suppose
for contradiction that this is not the case. Consider any planar drawing $\Pi$
of $K$. Then $\Pi(C)$ is a simple closed curve in the plane, and without loss
of generality we can assume that $\Pi(I)$ is entirely contained in the
interior of $\mathbb R^2\setminus\Pi(C)$. Now we define a new drawing
$\Delta'$ of $G$ that is identical with $\Delta$ on $G\setminus I$ and
homeomorphic to $\Pi$ on $K$. Since none of the edges in $F_i$ is involved in
any crossing of $\Delta$, this can be done in such a way that none of the
edges in $F_i$ is involved in any crossing of $\Delta'$. But then the number
of crossings of $\Delta'$ is smaller than that of $\Delta$, because the
restricion of $\Delta'$ to $K$ is planar. This contradicts the minimality of
the crossing number of $\Delta$.

Hence the restriction of $\Delta$ to $K$ is planar. In particular, this means
that none of the edges of $F_2$ is involved in any crossing of $\Delta$. By
the minimality of $i_0$, this implies $i_0=2$. \emph{Thus, surprisingly, $i_0$
  is independent of the drawing $\Delta$.}

\begin{figure}[ht]
\centering
\includegraphics[width=12cm]{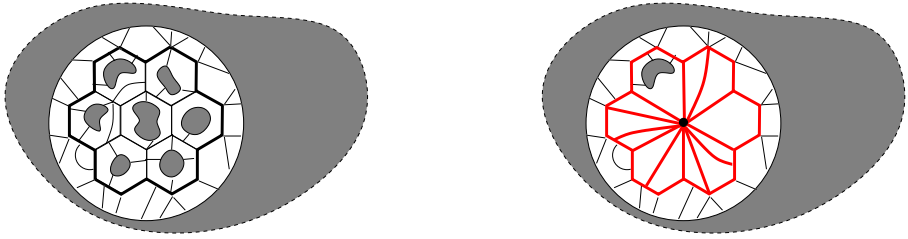}
\caption{The transformation from a graph $G$ to $G'$}\label{fig:contract}
\end{figure}

Let $G'$ be the graph obtained from $G$ by
contracting the connected subgraph $I$ to a single vertex $v_I$ (see
Figure \ref{fig:contract}).\footnote{In
  other words, $G'$ is obtained from $G$ by deleting all vertices of $I$,
  deleting all edges with both endpoints in $I$, adding a new vertex $v_I$,
  and replacing, for all edges with one endpoint in $I$, this endpoint by
  $v_I$.}  

Let $F'$ be the union of $F$ with the set of all edges of $h(C)$ and all edges
incident with the new vertex $v_I$.  \emph{Then $G$ has a $k$-good drawing
  with respect to $F$ if, and only if, $G'$ has a $k$-good drawing with
  respect to $F'$.} The forward direction of this claim is obvious by the
construction of $G'$ and $F'$, and for the backward direction we observe that
every $k$-good drawing $\Delta'$ of $G'$ with respect to $F'$ can be turned
into a $k$-good drawing of $G$ with respect to $F$ by embedding the planar
graph $I$ into a small neighborhood of $\Delta'(v_I)$.

Clearly, given $G,F$ and $h$, the graph $G'$ and the edge-set $F'$ can be
computed in linear time. Moreover $|V^{G'}|<|V^G|$. Combining this with Lemma
\ref{lem:ml1}, we obtain:

\begin{lem}\label{lem:ml2}
  There is a linear time algorithm that, given a graph $G$, either recognizes
  that the crossing number of $G$ is greater than $k$ or recognizes that the
  tree-width of $G$ is at most $w$ or computes a graph $G'$ and an edge set
  $F'\subseteq E^{G'}$ with $|V^{G'}|<|V^G|$ such that $G$ has a $k$-good drawing
  with respect to $F$ if, and only if, $G'$ has a $k$-good drawing with respect to
  $F'$.
\end{lem}

Iterating the algorithm of the lemma, we obtain:

\begin{cor}
  There is a quadratic time algorithm that, given a graph $G$, either
  recognizes that the crossing number of $G$ is greater than $k$ or computes a
  graph $G'$ and an edge set $F'\subseteq E^{G'}$ such that the tree-width of
  $G'$ is at most $w$ and $G$ has a $k$-good drawing with respect to $F$ if, and
  only if, $G'$ has a $k$-good drawing with respect to $F'$.
\end{cor}

\subsection*{Phase II}
If the algorithm has not found out that the graph has crossing number greater
than $k$ in Phase I, it has produced a graph $G'$ of tree-width at most $w$
and a set $F'\subseteq E^{G'}$ such that $G$ has a $k$-good drawing with
respect to $F$ if, and only if, $G'$ has a $k$-good drawing with respect to
$F'$. In Phase II, the algorithm has to decide whether $G'$ has a $k$-good
drawing with respect to $F'$.  Using Courcelle's Theorem \ref{theo:cou}, we
prove that this can be done in linear time.

To this end, we shall find an MSO-formula $\phi(X)$ such that for every graph
$G$ and every set $F\subseteq E^G$ we have $G\models\phi(F)$ if, and only if,
$G$ has a $k$-good drawing with respect to $F$. We rely on the well-known fact
that there is an MSO-formula $\phi_{\text{planar}}$ saying that a graph is
planar. (Actually, this is quite easy to see: $\phi_{\text{planar}}$ just says
that $G$ neither contains $K_5$ nor $K_{3,3}$ as a topological subgraph. Also see \cite{cou00}.)

\begin{figure}[ht]
\centering
\includegraphics[height=2.5cm]{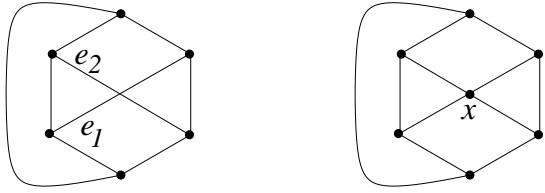}
\caption{A graph $G$ with selected edges $e_1,e_2$ and the resulting
  $G^{e_1\times e_2}$}\label{fig:ge}
\end{figure}

For a graph $G$ and distinct edges $e_1,e_2\in E^G$ we let $G^{e_1\times e_2}$
be the graph obtained from $G$ by deleting the edges $e_1$ and $e_2$ and
adding a new vertex $x$ and four edges connecting $x$ with the endpoints of
the edges of $e_{1}$ and $e_{2}$ in $G$ (see Figure \ref{fig:ge}).\label{Ge}
Observe that for every $l\ge 1$ a graph $G$ has an $l$-good drawing with
respect to an edge set $F\subseteq E^G$ if, and only if, there are distinct
edges $e_1,e_2\in E^G\setminus F$ such that $G^{e_1\times e_2}$ has an
$(l-1)$-good drawing with respect to $F$.

A standard technique from logic, the method of syntactical interpretations,
(easily) yields the following lemma:\footnote{For an introduction to the
  technique we refer the reader to \cite{ebbflutho94}, for the
  particular situation of MSO on graphs to \cite{cos93,cou97}.}

\begin{lem}
For every MSO-formula $\phi(Y)$ there exists an MSO-formula
$\phi^*(x_1,x_2,Y)$ such that for all graphs $G$, edge sets $F\subseteq E^G$
and  distinct edges $e_1,e_2\in E^G\setminus F$ we have:
\[
G\models\phi^*(e_1,e_2,F)\iff G^{e_1\times e_2}\models\phi(F).
\]
\end{lem}

Using this lemma, we inductively define, for every $l\ge 1$, formulas
$\phi_l(Y)$ and $\psi_l(x_1,x_2,Y)$ such that for every graph $G$ and edge set
$F\subseteq E^G$ we have
\[
G\models\phi_l(F)\iff\text{$G$ has an $l$-good drawing with respect to $F$},
\]
and for all $G$, $F\subseteq E^G$, and $e_1,e_2\in E^G\setminus F$ we have
\[
G\models\psi_l(e_1,e_2,F)\iff G^{e_1\times e_2}\text{ has an $(l-1)$-good
  drawing with respect to $F$}.
\]
We let 
\[
\psi_1(x_1,x_2,Y):=\phi_{\text{planar}}^*(x_1,x_2)
\]
and, for $l\ge 1$,
\begin{align*}
\phi_l(Y):=&\exists x_1\exists x_2\big(x_1\neq x_2 \wedge Ex_1\wedge Ex_2\wedge\neg Yx_1\wedge\neg
Yx_2\wedge\psi_{l}(x_1,x_2,Y)\big),\\
\psi_{l+1}(x_1,x_2,Y):=&\phi_l^*(x_1,x_2,Y).
\end{align*}
This completes our proof.

\subsection*{Computing a Good Drawing}
So far we have only proved that there is a quadratic time algorithm deciding if a graph $G$ has a good drawing with respect to a set $F\subseteq
E^G$. 

It is not hard to modify the algorithm so that it actually computes a drawing:
For Phase I, we observe that if we have a good drawing of $G'$ with respect
to $F'$ then we can easily construct a good drawing of $G$ with respect to
$F$. So we only have to worry about Phase II. 

By induction on $l$, for every $l\ge 0$ we define a linear-time procedure
$\text{DRAW}_l$ that, given a graph $G$ of tree-width at most $w$ and a subset
$F\subseteq E^G$, computes an $l$-good drawing of $G$ with respect to $F$ (if
there exists one). $\text{DRAW}_0$ just has to compute a planar drawing of
$G$.

For $l\ge 1$, we apply Theorem \ref{theo:cou2} to the
MSO-formula 
\[
\chi_l(x_1,x_2,Y):=x_1\neq x_2\wedge Ex_1\wedge Ex_2\wedge\neg Yx_1\wedge\neg
Yx_2\wedge\psi_l(x_1,x_2,Y).
\]
It yields a linear time algorithm that, given a graph $G$ and an $F\subseteq
E^G$, computes two edges $e_1,e_2\in E^G\setminus F$ such that
$G\models\chi_l(e_1,e_2,F)$ (if such edges exist). It follows immediately from
the definition of $\psi_l$ that
$G\models\chi_l(e_1,e_2,F)$ if, and only if, $G^{e_1\times e_2}$ has an $l$-good
drawing with respect to $F$.

Given $G$ and $F$, the procedure $\text{DRAW}_l$ applies this linear-time
algorithm to compute $e_1,e_2$ such that $G\models\chi_l(e_1,e_2,F)$. Then it
applies $\text{DRAW}_{l-1}$ to the graph $G^{e_1\times e_2}$ to compute an
$(l-1)$-good drawing of a graph $G^{e_1\times e_2}$ with respect to $F$. It modifies this drawing in a straightforward way to obtain an $l$-good
drawing of $G$ with respect to $F$.

\subsection*{Avoiding Logic}
For those readers who are not so fond of logic, let me briefly sketch
how the use of Courcelle's Theorem can be avoided. We have to find an
algorithm that, given a graph $G$ of tree-width at most $w$ and a set
$F\subseteq E^G$, decides whether $G$ has a good drawing with respect
to $F$. 

Let $l\ge 1$. For a graph $G$ and pairwise distinct edges
$e_1,\ldots,e_{2l}\in E^G$ we let
\[
G^{\times\bar e}:=\big(\cdots\big((G^{e_1\times e_2})^{e_3\times
  e_4}\big)\cdots\big)^{e_{l-1}\times e_{l}},
\]
that is, the graph obtained from $G$ by crossing $e_1$ with $e_2$, $e_3$ with
$e_4$, et cetera. Observe that, for every graph $G$, there exist an $l\le k$
and pairwise distinct edges
$e_1,\ldots,e_{2l}\in E^G$ such that $G^{\times\bar e}$ is planar if, and only
if, $G$ has a drawing with at most $k$ crossings such that every edge of $G$
is involved in at most one crossing of this drawing. This is not the same as
saying that the crossing number of $G$ is at most $k$.

However, there is a simple trick that makes it possible to work with
$G^{\times\bar e}$ anyway: For every graph $G$ we let $\tilde G$ be the graph
obtained from $G$ by subdividing every edge $(k-1)$-times, that is, by
replacing every edge by a path of length $k$. For $F\subseteq E^G$, we let
$\tilde F$ be the set of all edges of $\tilde G$ that appear in a subdivision
of an edge in $F$. Then clearly, $G$ has a $k$-good drawing with respect to
$F$ if, and only if, $\tilde G$ has a $k$-good drawing with respect to $\tilde
F$. The crucial observation is that $\tilde G$ has a $k$-good drawing with
respect to $\tilde F$ if, and only if, there exists an $l\le k$ and pairwise distinct edges
$e_1,\ldots,e_{2l}\in E^{\tilde G}\setminus\tilde F$ such that $\tilde
G^{\times\bar e}$ is planar. Note, furthermore, that the pair $(\tilde
G,\tilde F)$ can be constructed from $(G,F)$ in linear time.

Thus it suffices to find for every $l\ge 1$ a linear time algorithm that,
given a graph $G$ of tree-width at most $w$ and a set $F\subseteq E^G$,
computes pairwise distinct edges $e_1,\ldots,e_{2l}\in E^G\setminus F$ such
that $G^{\times\bar e}$ is planar (if such edges exist). 

Our algorithm first computes a tree-decomposition of $G$ of width at most $w$
using Bodlaender's linear time algorithm \cite{bod96}. Then by the usual
dynamic programming technique on tree-decompositions of graphs it computes
edges $e_1,\ldots,e_{2l}\in E^G\setminus F$ such that the graph
$G^{\times\bar e}$ neither contains $K_{3,3}$ nor $K_5$ as a topological
subgraph. By Kuratowski's Theorem, this is equivalent to $G^{\times\bar e}$
being planar.

The advantage of our approach using definability in monadic
second-order logic is that we have a precise proof without working out
the tedious details of what is sloppily described as the ``usual dynamic
programming technique'' above. 

\subsection*{Uniformity}
Inspection of our proofs and the proofs of the results we used shows
that actually there is \emph{one} algorithm that, given a graph $G$
with $n$ vertices and a non-negative integer $k$, decides whether the
crossing number of $G$ is at most $k$ in time $O(f(k)\cdot n^2)$ for a
suitable function $f$. Furthermore, it can be proved that $f$ can be
chosen to be of the form $2^{2^{p(k)}}$ for a polynomial $p$.

\section{Conclusions}
We have proved that for every $k\ge 0$ there is a quadratic time
  algorithm deciding whether a given graph has crossing number at most
  $k$. The running time of our algorithm in terms of $k$ is
  enormous, which makes the algorithm useless for practical
  purposes. This is partly due to the fact that the algorithm heavily relies on graph minor theory.
 
However, knowing the crossing number problem to be
fixed-parameter tractable may help to find better algorithms that are
practically applicable for small values of $k$. This has happened in a
similar situation for the vertex cover problem. The first proof
\cite{fellan88} that vertex cover is fixed-parameter tractable used
Robertson and Seymour's theorem that classes of graphs closed under
taking minors are recognizable in cubic time. Starting from there,
much better algorithms have been developed; by now, vertex cover can be
(practically) solved for a quite reasonable problem size (see
\cite{felste99} for a state-of-the-art algorithm).

%\bibliographystyle{plain}
%\bibliography{endlmod,grohe}

\pagebreak
\rnc{\thesection}{A}
\setcounter{theo}{0}
\section*{Appendix A: Monadic Second Order Logic}
We first
explain the syntax of MSO: We have an
infinite supply of \emph{individual variables}, denoted by $x,y,z,x_1$ et
cetera, and also  an infinite supply of set variables, denoted by $X,Y$, et
cetera. \emph{Atomic MSO-formulas (over graphs)} are formulas of the form $Vx$, $Ex$, $Ixy$, and
$Xx$, where $x,y$ are individual variables and $X$ is a set variable. The
class of MSO-formulas is defined by the following rules:
\begin{itemize}
\item
Atomic MSO-formulas are MSO-formulas.
\item
If $\phi$ is an MSO-formula, then so is $\neg\phi$. 
\item
If $\phi$ and $\psi$ are MSO-formulas, then so are $\phi\wedge\psi$,
$\phi\vee\psi$, and $\phi\to\psi$.
\item
If $\phi$ is an MSO-formula and $v$ is a variable (either an individual
variable or a set variable), then $\exists
v\phi$ and $\forall v\phi$ are MSO-formulas.
\end{itemize}
Recall that $U^G=V^G\cup E^G$.
A \emph{$G$-assignment} is a mapping $\alpha$ that
associates an element of $U^G$ with every individual variable and a
subset of $U^G$ with every set variable. We inductively define what it
means that a graph $G$ together with an assignment $\alpha$ \emph{satisfies} an
MSO-formula $\phi$ (we write $(G,\alpha)\models\phi$):
\begin{itemize}
\item
$(G,\alpha)\models Vx\iff \alpha(x)\in V^G$,\\ 
$(G,\alpha)\models Ex\iff \alpha(x)\in E^G$,\\ 
$(G,\alpha)\models Ixy\iff \big(\alpha(x)\in V^G,\alpha(y)\in E^G,\alpha(x)\text{ endpoint of }\alpha(y)\big)$,\\ 
$(G,\alpha)\models Xx\iff
\alpha(x)\in\alpha(X)$,
\item
$(G,\alpha)\models\neg\phi\iff(G,\alpha)\not\models\phi$,
\item
$(G,\alpha)\models\phi\wedge\psi\iff\big((G,\alpha)\models\phi\text{ and
  }(G,\alpha)\models\psi\big)$,\\ 
and similarly for $\vee$, meaning ``or'', and
$\to$, meaning ``implies''.
\item
$(G,\alpha)\models\exists x\phi\iff\text{there exists an $a\in U^G$
  such that $(G,\alpha\frac{x}{a})\models\phi$}$, where $\alpha\frac{x}{a}$
denotes the assignment with $\alpha\frac{x}{a}(x)=a$ and
$\alpha\frac{x}{a}(v)=\alpha(v)$ for all $v\neq x$,\\
and similarly for
$\forall x$ meaning ``for all $a\in U^G$'',
\item
$(G,\alpha)\models\exists X\phi\iff\text{there exists an $A\subseteq U^G$
  such that $(G,\alpha\frac{X}{A})\models\phi$}$,\\ 
and similarly for
$\forall X$ meaning ``for all $A\subseteq U^G$''.
\end{itemize}
It is easy to see that the relation $(G,\alpha)\models\phi$ only depends on
the values of $\alpha$ at the \emph{free variables} of $\phi$, i.e.\ those
variables $v$ not occurring in the scope of a quantifier $\exists v$ or
$\forall v$. We
write $\phi(x_1,\ldots,x_k,X_1,\ldots,X_l)$ to denote that the free individual
variables of $\phi$ are among $x_1,\ldots,x_k$ and the free set variables are
among $X_1,\ldots,X_l$. Then for a graph $G$ and $a_1,\ldots,a_k\in U^G$,
$A_1,\ldots,A_l\subseteq  U^G$ we write
$G\models\phi(a_1,\ldots,a_k,A_1,\ldots,A_l)$ if for every assignment $\alpha$
with $\alpha(x_i)=a_i$ and $\alpha(X_j)=A_j$ we have $(G,\alpha)\models\phi$.
A \emph{sentence} is a formula without free variables.

For example, for the sentence 
\begin{align*}
\phi:=\exists X\exists Y\Big(&\forall x\big(Vx\to (Xx\vee Yx)\big)\\
\wedge&\forall x\forall
y\Big(\big(x\neq y\wedge\exists z(Ez\wedge Ixz\wedge Iyz)\big)\to\neg\big((Xx\wedge Xy)\vee(Yx\wedge Yy)\big)\Big)\Big)
\end{align*}
we have $G\models\phi$ if, and only if, $G$ is
2-colorable.

\end{document}